\documentstyle[12pt,twoside,fleqn,espcrc1,epsf]{article}
\newcommand{\npr}{$\Gamma_n/\Gamma_p \,$}
\newcommand{\ndec}{$\Gamma_n \,$}
\newcommand{\pdec}{$\Gamma_p \,$}
\begin{document}

\title{Decay of Hypernuclei}

\author{A. Parre\~no\thanks{Talk presented at the KEK-Tanashi 
International Symposium on Physics of Hadrons and Nuclei, Tokyo, Japan,
December 14-17, 1998. In honor of Prof. K. Yazaki.}
\address{Institute for Nuclear Theory, University of Washington, 
Box 351550\\
Seattle, WA 98195-1550},
C.\ Bennhold
\address{Center of Nuclear Studies, Department of Physics,\\ The George
         Washington University, Washington, D. C. 20052}
and
A.\ Ramos
\address{Departament d'Estructura i Constituents de la Mat\`eria,
Facultat de F\'{\i}sica,\\
Diagonal  647, E-08028 Barcelona}}

\date{\today}
\maketitle

\begin{abstract}

We present a nonrelativistic transition potential for the weak
strangeness-changing
reaction $\Lambda N \to NN$. The potential is based on a
one meson exchange model (OME),
where, in addition to the long-ranged pion, the exchange of
the pseudoscalar $K, \eta$, as well as
the vector $\rho, \omega, K^*$ mesons is considered. 
Results obtained for different hypernuclear decay observables are compared to
the available experimental data.

\end{abstract}

\section{Introduction}

As is well known, the mesonic decay ($\Lambda \to \pi N$) of a $\Lambda$ particle
in the nuclear medium is highly suppressed due to 
the Pauli blocking effect acting on the outgoing nucleon. 
In contrast, the nonmesonic (NM) decay mode
($\Lambda N \to NN$), where the mass difference between the initial
hyperon and nucleon is converted into kinetic energy for the outgoing nucleons,
is the dominant decay mode for medium to heavy hypernuclei.

During the last thirty years, many theoretical works have described the NM decay of
hypernuclei by mainly using two different approaches: a description in terms 
of quark degrees of freedom or/and the use of a OME model.
A recent review of the present status of the theoretical and
experimental situation can be found in Ref. \cite{RO98}.

The OME approach is the one that will be presented in this contribution.
In order to draw conclusions
regarding the weak dynamics, nuclear structure details have to be
treated with as few approximations and ambiguities as possible.
Working in a shell-model framework, spectroscopic factors are employed to describe 
the initial hypernuclear and final nuclear structure. 
To reduce the uncertainties regarding initial and final short-range correlations 
we use realistic $\Lambda N$ and $NN$ strong interactions based on the
Nijmegen baryon-baryon potential.
More details about this calculation can be found in Ref. \cite{PRB97}.

\section{The Meson Exchange Potential}

In a OME model, the transition $\Lambda N \to N N$ is assumed to proceed via
the exchange of virtual mesons belonging to the ground-state
pseudoscalar and vector meson octets. The nonrelativistic reduction
of the free space Feynman amplitude 
involving a weak and a strong baryon-baryon-meson (BBM) vertex, gives
the nonrelativistic potential in momentum space
\cite{PRB97}. The $\Delta I = 1/2$ rule, which is experimentally known to hold in the
free $\Lambda$ decay, has been assumed when computing the NM amplitudes. How to account 
for possible violations of this rule and their consequences is discussed below.
For pseudoscalar mesons, the transtion potential reads:
\begin{equation}
V_{ps}({\bf q}) = - G_F m_\pi^2
\frac{g}{2M} \left(
{\hat A} + \frac{{\hat B}}{2\overline{M}}
\mbox{\boldmath $\sigma$}_1 {\bf q} \right)
\frac{\mbox{\boldmath $\sigma$}_2 {\bf q} }{{\bf q}^2+\mu^2} \ ,
\label{eq:pspot}
\end{equation}
where $G_F m_\pi^2= 2.21\times 10^{-7}$ is the Fermi weak coupling constant, 
${\bf q}$ 
is the momentum carried by the meson, 
$g = g_{\rm {\scriptscriptstyle BBM} }$ the strong coupling constant for the 
BBM vertex, $\mu$ the meson mass,
$M$ the nucleon mass and $\overline M$
the average between the nucleon and $\Lambda$ masses.
The operators ${\hat A}$ and ${\hat B}$ contain the weak parity violating (PV)
and parity conserving (PC) coupling constants respectively, as well as the
isospin dependence of the potential.
For vector mesons the potential has the form:
\begin{eqnarray}
{V_{v}}({\bf q})  &=&
G_F m_\pi^2
 \left( F_1 {\hat \alpha} - \frac{({\hat \alpha} + {\hat \beta} )
 ( F_1 + F_2 )} {4M \overline{M}}
(\mbox{\boldmath $\sigma$}_1 \times {\bf q})
(\mbox{\boldmath $\sigma$}_2 \times {\bf q}) \right. \nonumber \\
& & \phantom { G_F m_\pi^2 A }
\left. - i \frac{{\hat \varepsilon} ( F_1 + F_2 )} {2M}
(\mbox{\boldmath $\sigma$}_1 \times
\mbox{\boldmath $\sigma$}_2 ) {\bf q}\right)
\frac{1}{{\bf q}^2 + \mu^2} \
\label{eq:rhopot}
\end{eqnarray}
with $F_1 = g^{\rm {\scriptscriptstyle V}}_{\rm{\scriptscriptstyle BBM}}$ and
$F_2 = g^{\rm {\scriptscriptstyle T}}_{\rm {\scriptscriptstyle BBM}}$
the vector and tensor strong couplings.
The (PC) ${\hat \alpha}$, ${\hat
\beta}$ and (PV) ${\hat \varepsilon}$ operators
contain the isospin structure together with the corresponding weak couplings.
In the case of isovector mesons ($\pi$,$\rho$) the
isospin factor is $\mbox{\boldmath $\tau$}_1
\mbox{\boldmath $\tau$}_2$, and for isoscalar mesons ($\eta$,$\omega$)
this factor is just $\hat{1}$ for all spin structure pieces of the potential.
In the case of isodoublet mesons ($K,K^*$) there
are contributions proportional to $\hat{1}$ and to $\mbox{\boldmath
$\tau$}_1 \mbox{\boldmath $\tau$}_2$ that depend on the
coupling constants and, therefore, on the
spin structure piece of the potential.

In order to account for the finite size and structure of baryons and mesons, a monopole 
form factor $F({\bf q}^2)= (\Lambda^2-\mu^2)/(\Lambda^2+{\bf q} ^2)$
is considered in both the strong and weak vertices, where
the value of
the cut-off, $\Lambda$, different for each meson, is taken
from the J\"ulich $YN$ interaction\cite{juelich}. 
To incorporate the effects of the $NN$ interaction, we solve a T-matrix scattering
equation in momentum space, which describes the relative motion of two nucleons moving 
under the influence of 
the strong interaction. For the strong $NN$ interaction we use 
the updated version of the Nijmegen $NN$ potential\cite{stoks}.
For the bound $\Lambda N$ state, we replace the uncorrelated shell-model
$\Lambda N$ wave function
(for which we take harmonic oscillator solutions) by a correlated
$\Lambda N$ wave function that contains the effect of the
strong $\Lambda N$ interaction.
This wave function is obtained by taking, as a guide,
microscopic finite-nucleus $G$-matrix
calculations\cite{halder} using the soft-core and hard-core
Nijmegen models\cite{nijme}. It has been shown \cite{sitges} 
that multiplying the uncorrelated
$\Lambda N$ wave function with the spin-independent
correlation function
\begin{equation}
f(r)=\left( 1 - {\rm e}^{-r^2 / a^2} \right)^n + b r^2 {\rm
e}^{-r^2 / c^2} \ ,
\label{eq:cor}
\end{equation}
with $a=0.5 $, $b=0.25 $, $c= 1.28$, $n= 2$,
yields decay rates which lie in between 
those using the hard and soft Nijmegen $\Lambda N$ potential models.
Since the
deviations were at most 10\% the above parametrization can be used as a
good approximation to the full correlation function.

The strong BBM couplings are taken from either the Nijmegen\cite{nijme} or 
the J\"ulich \cite{juelich} potentials.
In the weak sector, only the couplings corresponding to decays
involving pions are experimentally known. The weak couplings
involving mesons heavier than the pion are obtained following Refs.
\cite{holstein,delatorre}.
The PV amplitudes for 
the nonleptonic decays $B \rightarrow B' + M$ involving pseudoscalar mesons
are derived using the soft-meson reduction theorem and SU(3) symmetry, which
allows us to 
relate the physical amplitudes of the nonleptonic hyperon decays
into a pion plus a nucleon or a hyperon, $B \rightarrow B' + \pi$, with
the unphysical amplitudes of the other members of the meson octet
($K, \eta$). On the other hand, SU(6$)_w$ permits
relating the amplitudes involving pseudoscalar mesons with those of the
vector mesons.
The pole model, where the pole terms due to the $(\frac{1}{2})^+$ ground
state (singular in the SU(3) soft meson limit) are assumed to be the dominant 
contribution, is used for obtaining the PC coupling constants. 

\section{Results}

In Table \ref{tab:rate} we explore the effect of including
all the mesons on the weak decay observables
for the $^{12}_\Lambda $C hypernucleus. These observables include: the 
NM decay rate in units of the free $\Lambda$ decay rate, 
$\Gamma_\Lambda = 3.8 \times 10^9 s^{-1}$ (second column), 
the fraction between the neutron
($\Lambda n \to nn$) and the proton ($\Lambda p \to np$) induced decays
(third column) and the intrinsic $\Lambda$ asymmetry parameter, $a_\Lambda$,
characteristic of the elementary $\vec{\Lambda} N \to NN$ reaction taking place 
in the nuclear medium (last column). 
This last parameter is related to the observed angular asymmetry in the distributions of 
protons coming from the decay of polarized hypernuclei.
The numbers between parenthesis are obtained when the J\"ulich
constants are used in the strong vertex instead of the Nijmegen ones.
Note that, through the pole model, this choice of strong coupling constants 
would also affect the couplings in the PC weak vertices. However, we keep them 
fixed to the values obtained using the Nijmegen model and modify only the 
strong vertex. This allows us to assess the effect of using different sets of 
strong couplings derived from $YN$ potentials that fit the hyperon-nucleon 
scattering data equally well.

Even if not shown here, the pion-exchange (OPE) contribution dominates 
the rate not only in magnitude but in range,
a consequence of the pion being the lightest meson. 
This rate is especially sensitive to the inclusion of the strange
mesons, while including the $\rho$-meson has almost no effect.
Note that the contribution of each isospin-like pair
[$(\pi,\rho)$, ($K,K^*$), $(\eta,\omega)$] interferes destructively,
so, the reduction caused by the $K$ meson is mostly 
compensated for by adding the $K^*$. A similar situation is observed between
the $\eta$ and $\omega$ mesons, consequently their combined effect on the
rate
is negligible. The final result for the rate is 15\% smaller or greater
than the
pion-only one, depending on the choice of couplings in the strong sector.
This sensitivity is unfortunate since
it will certainly complicate the task of extracting weak couplings from
this reaction. Improved $YN$ potentials which narrow the range of the
strong coupling constants are required to reduce this uncertainty.
Table \ref{tab:rate} also shows the results obtained when the $NNK$
weak coupling constants derived with one-loop corrections to the leading
order in $\chi$PT\cite{spring} are used (last row). Due to the smaller value of the
coupling constants the effect of the $K$ meson is reduced and thus
the total rate is increased by about 10\%.

It has been known for a long time that the large tensor transition 
($^3S_1 \to ^3D_1$) in the OPE mechanism, where only $T=0$ $np$ pairs can occur,
is the reason for the small value of \npr given by 
pion exchange alone\cite{mckellar}.
For many years, it was
believed that the inclusion of
additional mesons would dramatically increase this observable.
But here, we find the opposite to be true.
This ratio is, as
expected, quite sensitive to the isospin structure of the exchanged
mesons, and it is again the inclusion of the two strange
mesons that dramatically modifies this partial ratio.
Including the $K$-exchange which interferes
destructively with the pion amplitude in the neutron-induced channel
leads to a reduction
of the ratio by more than a factor of three.
The $K^*$, on the other hand, adds contructively.
Using the Nijmegen strong couplings constants leads to a final ratio that is
34\% smaller than the pion-only ratio, while using
the J\"ulich couplings leaves this ratio unchanged,
due mostly to the larger $K^*$ and $\omega$
couplings.
Employing the weak $NNK$ couplings calculated with $\chi$PT
increases the  $\Gamma_n/\Gamma_p$ ratio by 17\% with
Nijmegen couplings while the ratio remains
unchanged for the J\"ulich model.

The intrinsic asymmetry parameter, $a_\Lambda$, 
is also found to be very sensitive to the different
mesons included in the model. This is the only observable which is
changed dramatically by the inclusion of the $\rho$, reducing the
pion-only value by more than a factor of two. Adding the other mesons
increases $a_\Lambda$, leading to a result about 30\% or 50\%
larger than for $\pi$-exchange alone, again depending on the type of strong
couplings used.

Our final results for various hypernuclei are compared with experimental
data in Table \ref{tab:hyps}. We find overall agreement
between our results
for the nonmesonic rate and the experimental values, especially when
the $\chi$PT weak couplings for the $K$ meson are used.
The situation for the \npr ratio is different, and
the theoretical values greatly underestimate the newer
central experimental ones, although the large experimental error
bars do not permit any definite conclusions at this time.
On the other hand, the proton-induced rate which has errors of the same
magnitude as the total rate is overpredicted by our calculations by up
to a factor of two. Even though \ndec and \pdec 
appear in disagreement with the data, their sum
conspires to give a total rate which reproduces the measurements.

Motivated by the hope that $\Delta I=1/2$ violations would
affect the \npr ratio, we studied the effect of allowing $\Delta I=3/2$
transitions in the vector meson exchange potential\cite{PLB98}.
The new weak couplings were derived in the factorization approximation
and, in order to take into account the
limitations imposed by this derivation,
we allowed them to vary by up to a factor
of $\pm 3$. While the total decay rate changed by at most 6\%,
the \npr ratio could be enhanced by a factor of 2, for different combinations of the 
multiplying factors. These modifications were almost due to changes 
in \ndec; \pdec was barely afected. Even though the estimates based
on the factorization approximation are rather crude, the variation on the
results for the \npr ratio and for the asymmetry parameter (which can be altered by a
factor of 7) clearly indicates that one cannot assume
the validity of the $\Delta I=1/2$ rule. Experiments measuring partial 
decay rates of light 
hypernuclei are desirable in order to clarify the validity of such rule.

Regarding the asymmetry parameter,
comparison with experiment can only be made at the level of the
measured proton asymmetry, which is determined as a product of
$a_\Lambda$, characteristic of the weak
decay process, and the polarization of the $\Lambda$ inside the hypernucleus, 
$p_\Lambda$,characteristic of the strong formation process and 
which must be determined theoretically.
In order to avoid the need for theoretical input and access $A_p$
directly, a new experiment at KEK has measured
the decay of polarized $^{5}_\Lambda$He extracting 
the pion asymmetry from the mesonic
channel, ${\cal A}_{\pi^-}$\cite{ajim98}.
The asymmetry parameter $a_{\pi^-}$ of the pionic channel has
been estimated to be very similar to that of the free $\Lambda$
decay\cite{motoba}, and, therefore, the hypernuclear polarization is
obtained from the relation $P_y={\cal A}_{\pi^-}/a_{\pi^-}$.
This value, together with the measurement of the proton asymmetry from the
nonmesonic decay has allowed to get the value of $a_\Lambda$ for 
$^{5}_\Lambda$He, which has been quoted as $0.22
\pm 0.20$\cite{ajim99}, in disagreement with the present theoretical
predictions.

\section{Summary and Outlook}

Total decay rates evaluated with the full weak OME potential fall within
15\% of the value obtained with pion exchange only and reproduce the
experimental data. This is partly due to the destructive interference
between the contributions of the heavier mesons whose individual influence on the
decay rate can be substantial. 
The importance of kaon exchange makes it
possible to see the effects of modifying the weak $NNK$ couplings by
one-loop corrections to the leading order in
$\chi$PT. Including these loop
graphs leads to a reduction of the $NNK$ couplings from their tree-level
value up to 50\%, which in turn modifies the decay rate by up to
20\%. Future experiments should be able to verify this effect.
We found the influence 
of strange mesons to be even more pronounced in
the partial rates and their ratio.
Furthermore, this ratio turns out to be
sensitive to the choice of strong coupling constants as well.
This finding indicates the need for improved $YN$
potentials with better determined strong couplings at the
hyperon-nucleon-meson vertices.
In contrast to the previous observables we found the proton asymmetry to
be very sensitive to the $\rho$-exchange while the influence of the
kaon is more moderate. This polarization observable is therefore an
important addition to the set of observables since its sensitivities are
different from the total and partial rates.

Within the one-meson exchange picture it would be desirable to use weak
coupling constants obtained from more sophisticated approaches. A
beginning has been made by Savage and Springer\cite{spring}
however, an understanding of the weak
$\Lambda {\rm N} \pi$ and $\Sigma {\rm N} \pi$ couplings within the
framework
of chiral lagrangians is still missing. Furthermore, due to the
importance of the $K^*$-meson it would be desirable to recalculate its
weak $NNK^*$ couplings in improved models as well.

In an attempt to solve the \npr puzzle, 
the authors of Ref.\cite{ramos4}
studied the 3N emission channel ($\Lambda N N \to N N N$),
where the virtual pion emitted at the weak vertex is
absorbed by a pair of nucleons which are correlated through the strong
force,
including final state interactions of the three nucleons on their
way out of the
nucleus via a Monte Carlo simulation. It was shown that the
new channel influences the analysis of the ratio \npr 
increasing its experimental error bars and leading to
a value
compatible with the predictions of the OPE model. However, the
same calculation shows that a comparison of the proton spectrum
with the experimental one favors values of $\Gamma_n/\Gamma_p=$2--3.
It is therefore
imperative, before speculating further about the deficiencies
of the present
models in reproducing this ratio, to carry out more precise
experiments
such as the measurement of the number of protons emitted per
$\Lambda$ decay.

We have seen that the consideration of the exchange of mesons heavier than the
pion in the NM decay of hypernuclei does not change the total and partial 
rates in a dramatic way. A different approach to the problem can be found in
Ref.\cite{OTIS98}, where the authors compare the Direct Quark (DQ) potential with a 
conventional OPE one, and calculate the NM decay rates of light hypernuclei. 
In order to fix the relative phase between both contributions, the weak 
$\Lambda N \pi$ coupling is related to a baryon matrix element of the weak hamiltonian
for quarks by using soft-pion relations. They found that the DQ contributes the most 
in $J=0$ transitions, enhancing \ndec and therefore the neutron-to-proton ratio, and 
that the $\Delta I =3/2$ components are significant in the $J=0$ transitions. 
On the other hand, their nuclear matter results show that a softer cut-off for the 
pion ($\approx$ 800 MeV) compared to the Born-like one used in the present contribution 
(1.3 GeV), seems more appropriate to reproduce experimental values of \pdec\cite{Ikek98}. 

The need for improved $YN$ interaction models has also been pointed out
in Ref. \cite{PRKB98}. Motivated by future measurements
\cite{kishi98} of cross sections and different polarization
observables \cite{kishi99} for the $pn \to p \Lambda$
reaction near threshold, Ref. \cite{PRKB98} gives a theoretical
prediction using the same model as presented here, and taking
advantage of the lack of nuclear structure ingredients. This study has shown
that this reaction is not only sensitive to the weak ingredients of the model,
but also to the strong $YN$ interaction. 
The cross sections have been found to be of the order of $10^{-12} mb$, a good 
challenge for the improved experimental devices.
The experiment should shed some light onto the understanding of the
weak $\Delta S = 1$ hadronic interaction.

The hypernuclear weak decay studies are being extended to
double-$\Lambda$ hypernuclei\cite{PRBfutur}. Very few events involving
these exotic objects --- whose
very existence would place stringent constraints on the existence of the
elusive H-dibaryon --- have been reported. Studying the weak decay of
these objects would open the door to a number of new exotic
$\Lambda$-induced decays: $\Lambda \Lambda \to \Lambda N$ and $\Lambda
\Lambda \to \Sigma N$. Both of these decays would involve hyperons in
the final state and should be distinguishable from the ordinary $\Lambda N \to
 NN$ mode.  Especially the $\Lambda \Lambda \to \Lambda N$ channel would be
intruiging since the dominant pion exchange is forbidden, thus this
reaction would have to occur mostly through kaon exchange.  One would
therefore gain access to the weak $\Lambda \Lambda K$ vertex.

The promising hypernuclear program at KEK, after an improved measurement of
the $^{5}_\Lambda$He decay, the continuing program at BNL, which recently
proved the existence of the $^4_\Sigma$He hypernucleus, and the hypernuclear 
physics program (FINUDA) at DA$\Phi$NE, 
represent excellent opportunities to obtain new valuable information
that will shed light onto the
still unresolved problems of the weak decay of hypernuclei.

\begin{table}
\centering
\caption{Nonmesonic decay observables of
 ${}^{12}_\Lambda$C. The values
in parentheses have been calculated using the J\"ulich-B coupling
constants at the strong vertex.}
\vskip 0.1 in
\begin{tabular}{lccc}
 &$\Gamma^{nm}/\Gamma_\Lambda$ & $\Gamma_n/\Gamma_p$ & $a_\Lambda$ \\
\hline\hline
$\pi$ & 0.89 (0.89) & 0.10 (0.10) & $-0.24$ $(-0.24)$ \\
$+ \rho$ & 0.86 (0.83)& 0.10 (0.10) & $-0.10$ $(-0.05)$ \\
$+ K$ & 0.50 (0.51)& 0.03 (0.03) & $-0.14$ $(-0.07)$ \\
$+ K^*$ & 0.76 (0.90) & 0.05 (0.07) & $-0.18$ $(-0.20)$\\
$+ \eta$ & 0.68 (0.90)  & 0.06 (0.07) & $-0.20$ $(-0.20)$\\
$+ \omega$ & 0.75 (1.02) & 0.07 (0.11) & $-0.32$ $(-0.37)$\\
\hline
 &  & &  \\
\parbox{3cm}{weak $K$-couplings from $\chi$PT \cite{spring}} &
  0.84 (1.10)& 0.08 (0.11) & $-0.30$ $(-0.35)$ \\
 &  & &  \\ 
\end{tabular}
\label{tab:rate}
\end{table}

\begin{table}
\centering
\caption{Weak decay observables for various hypernuclei.
The values in parentheses have been calculated using the
$NNK$ weak coupling constants obtained when including one-loop corrections
to the leading order in
$\chi$PT\protect\cite{spring}.}
\vskip 0.1 in
\begin{tabular}{lccc}
 & $^5_\Lambda$He
 & $^{11}_\Lambda$B
 & $^{12}_\Lambda$C \\
\hline
$\Gamma/\Gamma_\Lambda$ & 0.41 (0.47) & 0.61 (0.69) & 0.75
(0.84) \\
EXP: & $0.41\pm 0.14$ \cite{szymanski} & $0.95 \pm 0.13 \pm 0.04$ \cite{noumi}
 & $1.14\pm 0.2$ \cite{szymanski} \\
 & & & $0.89 \pm 0.15 \pm 0.03$
\cite{noumi} \\
\hline
$\Gamma_n/\Gamma_p$ & 0.07 (0.09) & 0.08 (0.10) & 0.07 (0.08) \\
EXP: & $0.93\pm 0.55$\cite{szymanski} &
       $1.04^{+0.59}_{-0.48}$ \cite{szymanski} & $1.33^{1.12}_{-0.81}$
       \cite{szymanski} \\
     &  & $2.16 \pm 0.58^{+0.45}_{-0.95}$ \cite{noumi} &
          $1.87 \pm 0.59^{+0.32}_{-1.00}$ \cite{noumi} \\
     &  & $0.70\pm 0.3$\cite{montwill} & $0.70\pm 0.3$\cite{montwill} \\
     &  & $0.52\pm 0.16$\cite{montwill} & $0.52\pm 0.16$\cite{montwill}\\
\hline
$\Gamma_p/\Gamma_\Lambda$ & 0.39 (0.43) & 0.56 (0.62) & 0.71 (0.78) \\
EXP: & $0.21 \pm 0.07$\cite{szymanski} & $0.30^{+0.15}_{-0.11}$\cite{noumi} &
$0.31^{+0.18}_{-0.11}$\cite{noumi} \\
\hline
$a_\Lambda$ & $-0.27$ ($-0.27$) & $-0.39$ ($-0.38$) &
              $-0.32$ ($-0.30$) \\
EXP: & 0.22 $\pm$ 0.20\cite{ajim99} & & \\
\hline
${\cal A}(0^{\rm o})$ &  &$-0.12$ ($-0.12$) & $-0.03$ ($-0.03$) \\
EXP: & & $-0.20\pm 0.10$ \cite{ajim98} & $-0.01\pm 0.10$ \cite{ajim98} \\
\end{tabular}
\label{tab:hyps}
\end{table}

\end{document}